\documentclass{phb-proc4-auth}
\usepackage{graphicx}
\usepackage{epsfig}

\usepackage{graphicx}
\usepackage{amssymb}

\begin{document}
\begin{frontmatter}


\journal{SCES '04}


\title{Gap Symmetry and Thermal Conductivity in Nodal Superconductors }

%
%
%
%
%
%

\author[ko]{H. Won\corauthref{hwadd}\thanksref{hw}
}
\author[us]{K. Maki}

%

\address[ko]{Department of Physics, Hallym University
Chuncheon 200-702, South Korea}
\address[us]{Department of Physics and Astronomy,
 University of Southern California, Los Angeles, CA 90089-0484,
USA}
%
%
%
%

\thanks[hw]{HW acknowledges the support from the
Korean Science and Engineering Foundation(KOSEF) through the Grant
No.R05-2004-000-1084}
%
%
%
%

\corauth[hwadd]{Corresponding Author: Dept. of Physics, Hallym
University, Chuncheon 200-702, South Korea.
Phone:(+82)-(33)248-2053, Fax:(+82)-(33) 254-2873,
Email:hkwon@hallym.ac.kr}


\begin{abstract}

Here we  consider the   universal heat conduction  and  the
angular dependent thermal conductivity in the vortex state for a
few nodal superconductors.  We present the thermal conductivity as
a function of impurity concentration and the angular dependent
thermal conductivity in a few nodal superconductors. This provides
further insight in the gap symmetry of superconductivity in
Sr$_2$RuO$_4$ and UPd$_2$Al$_3$.

\end{abstract}

%
%



\end{frontmatter}

%
%
%
%
%

Since the appearance  of heavy fermion  superconductors, organic
superconductors,  high T$_c$ cuprate superconductors,
Sr$_2$RuO$_4$, the gap symmetry has been  the central issue for
these new superconductors.   In the  last few  years, Izawa {\it
et al.} have succeeded  in identifying the  energy  gap  in
Sr$_2$RuO$_4$, CeCoIn$_5$, $\kappa$-(ET)$_2$Cu(NCS)$_2$,
YNi$_2$B$_2$C and PrOs$_4$Sb$_12$ \cite{1,2,3,4,5}. In these works
the angular dependent thermal conductivity in the vortex state of
nodal superconductors has played the crucial role \cite{6,7}. Here
we show two new  results: the universal heat conduction for
different nodal superconductors and the angular dependent thermal
conductivity for nodal superconductors with horizontal nodes.
Until now the universal heat  conduction is discussed only  for
d-wave superconductors in quasi  2D  systems (i.e. $\Delta({\bf
k}) =\Delta f  $ with $f=\cos(2 \phi)$ and $\sin(2 \phi)$)
\cite{7,8}. Here we consider in addition to $f$ for d-wave
superconductors  $f=\sin(\chi)$, $\cos(\chi)$, $\cos(2\chi)$,
$e^{\pm i \phi}\cos(\chi)$, and $e^{\pm i\phi}\sin\chi$ where
$\chi = c k_z$. Note all these states have the same quasiparticle
density of states, the gap equations and the
thermodynamics\cite{dahm}. Then we obtain
\begin{equation}\label{1}
\frac{\kappa_{xx}}{\kappa_n} = \frac2\pi \frac{\Gamma_{0}}\Delta
\frac{1}{\sqrt{1 +C_0^2}} E(\frac{1}{\sqrt{1 +C_0^2}}) \equiv
I_1(\frac{\Gamma}{\Gamma_0})
\end{equation}
where $\Gamma_0 \simeq 0.866T_c$, the quasiparticle scattering
rate at which the superconductivity disappears \cite{8} and $C_o$
is determined by
\begin{equation}\label{2}
\frac{C_0^2}{\sqrt{1+C_0^2}} K(\frac{1}{\sqrt{1+C_0^2}})= \frac
\pi2\frac{\Gamma}{\Delta}
\end{equation}
and $K(z)$  and $E(z)$  are  the complete  elliptic integrals.
Here we assume that the  impurity scattering is in the unitarity
limit. So we see that  Eq.(\ref{1}) is valid for all  $f$'s we
have described above.  In other words the  planar thermal
conductivity($\kappa_{xx}=\kappa_{yy}$) cannot discriminate
different nodal superconductors in the absence of magnetic field.
\par
On the other hand  $\kappa_{zz}$ is of more interest, we find
\begin{equation}\label{3}
  \frac{\kappa_{zz}}{\kappa_n} = I_1 ( \Gamma /\Gamma_0)
\end{equation}
for $f= \cos(2\phi)$, $\sin(2 \phi)$, and $\cos(2 \chi)$. But
\begin{eqnarray}\label{4}
  \frac{\kappa_{zz}}{\kappa_n} &=&
  \frac4\pi \frac{\Gamma_{0}}\Delta
\frac{1}{\sqrt{1 +C_0^2}} \big\{E(\frac{1}{\sqrt{1 +C_0^2}})
 -C_0^2 \big( \nonumber \\ & & \!\!\!\!\!\!\!\! K(\frac{1}{\sqrt{1 +C_0^2}})
-E(\frac{1}{\sqrt{1 +C_0^2}}) \big) \big\}
 \equiv
I_2(\frac{\Gamma}{\Gamma_0})
\end{eqnarray}
for $f= \cos(\chi)$, $e^{\pm i \phi}\cos(\chi)$ and
\begin{eqnarray}\label{5}
  \frac{\kappa_{zz}}{\kappa_n} &=& \frac{2\Gamma_{0}}\Delta \frac{\Gamma}\Delta
  \big\{1- E(\frac{1}{\sqrt{1 +C_0^2}})/
 K(\frac{1}{\sqrt{1
  +C_0^2}})\big\}\nonumber\\
  &=&I_3(\frac{\Gamma}{\Gamma_0})
\end{eqnarray}
for $f= \sin\chi$ and $e^{\pm i \phi}\sin\chi$. $I_1$, $I_2$ and
$I_3$ versus $\Gamma/\Gamma_0$ are shown in Fig. 1.
\par
Also the data by Suzuki {\it et al.} \cite{10} clearly favors
f-wave superconductor  $f= e^{\pm i \phi}\cos( \chi)$ as in
\cite{1}. Further recent magneto thermal conductivity data for
$\kappa$ in UPd$_2$Al$_3$ \cite{11} appear to be more consistent
with $f=\cos 2\chi$.
\begin{figure}[h]
\vspace{-2.5cm}
\centerline{\epsfig{figure=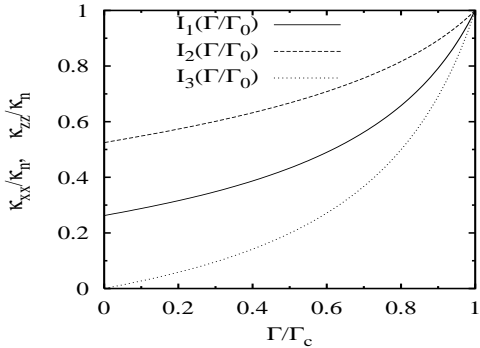,width=7cm,height=7.5cm}}
\caption{ $I_i(\Gamma/\Gamma_{c})$'s in Eq.(\ref{1}), Eq.(\ref{4})
 and Eq.(\ref{5})are shown,  which represent $\kappa_{xx}$ and $\kappa_{zz}$
 for various $f$'s. }
\end{figure}
\par
The angular dependent thermal conductivity in quasi 2D nodal
superconductors has been considered in \cite{12}. However in the
presence of magnetic field in the z-x plane has not been
considered. Following \cite{12} we obtain in the superclean limit
($\sqrt{\Gamma \Delta} \ll \tilde{v}\sqrt{eH} $ )
\begin{equation}\label{6}
  \kappa_{yy}/\kappa_n = \frac2\pi \frac{\tilde{v}^2{eH}
  }{\Delta^2}  F_1(\theta)
\end{equation}
and in the clean limit ($\tilde{v}\sqrt{eH} \ll
\sqrt{\Gamma\Delta} $ )
\begin{equation}\label{7}
 \kappa_{yy}/\kappa_{00} =1+  \frac{\tilde{v}^2{eH}
  }{6\pi\Gamma\Delta}  F_2(\theta) \ln(2
  \sqrt{\frac{2\Delta}{\pi\Gamma}})
   \ln(\frac{2\Delta}{\tilde{v}\sqrt{eH}})
\end{equation}
where
\begin{equation}\label{8}
  F_1(\theta)=\sqrt{\cos^2\theta + \alpha \sin^2\theta}\big(
  1+\sin^2\theta(-\frac38 + \alpha\sin^2\chi_0)\big)
\end{equation}
and
\begin{equation}\label{9}
 F_2(\theta)=\sqrt{\cos^2\theta + \alpha \sin^2\theta}\big(
  1+\sin^2\theta(-\frac14 + \alpha\sin^2\chi_0)\big)
\end{equation}
where $\alpha=(v_c/v_a)^2$ and $\chi_0$ is the nodal angle. For
example $f=\sin(\chi)$, $\cos(\chi)$ and $\cos(2\chi)$ gives
$\chi_0 =0$, $\frac\pi2$ and $\frac\pi4$, respectively. Here
$\kappa_n$ and $\kappa_{00}$ are the thermal conductivity in the
normal state and in the limit of $T \rightarrow 0$ respectively.
We show $F_1(\theta)$ and $F_2(\theta)$ for $\chi_0 =0$,
$\frac\pi4$ and $\frac\pi2$ in Fig.2. Here we took appropriate
$\alpha=0.69$ to UPd$_2$Al$_3$.
\begin{figure}[h]
\centerline{\epsfig{figure=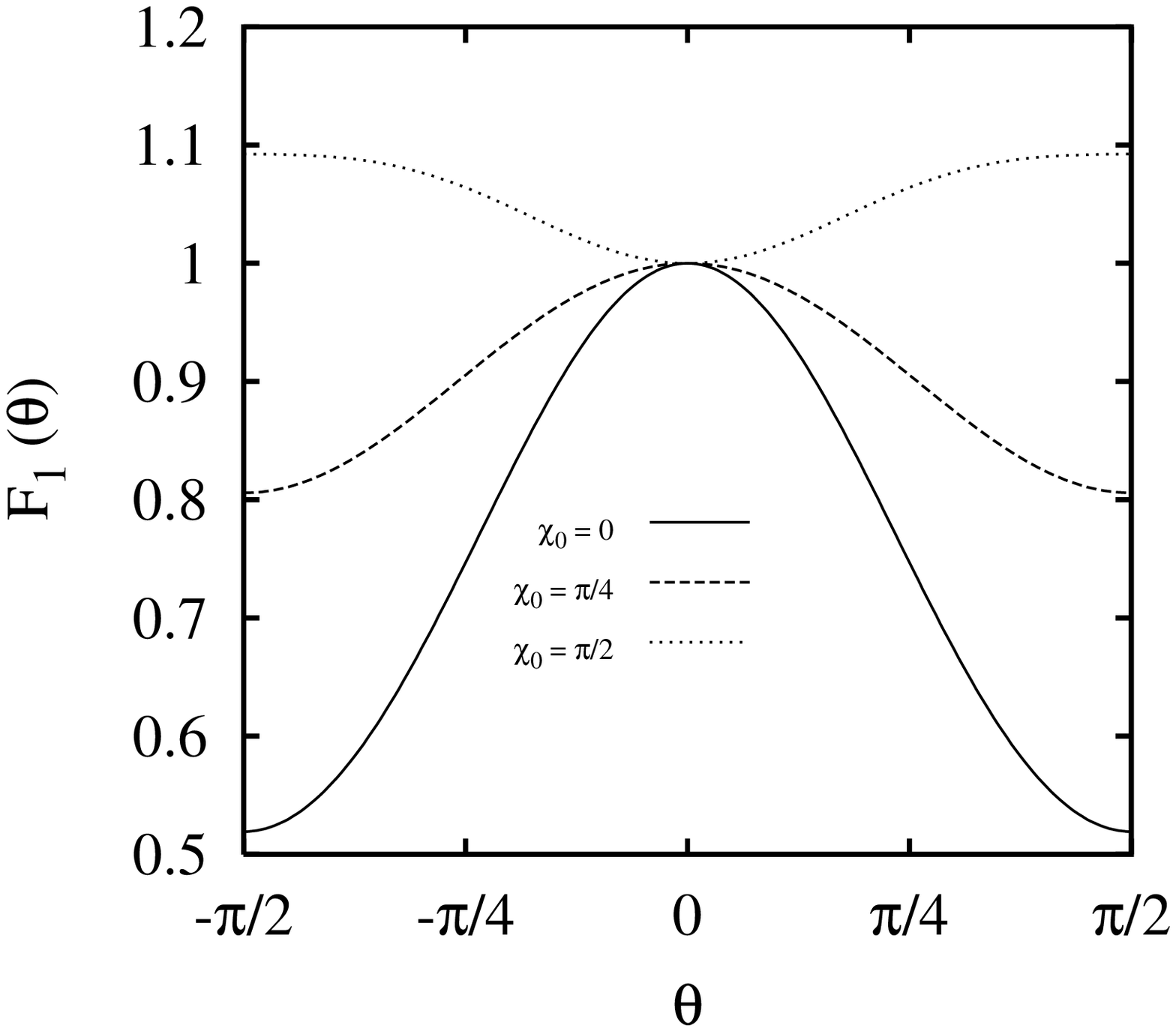,width=7cm,height=4.5cm}}
\centerline{\epsfig{figure=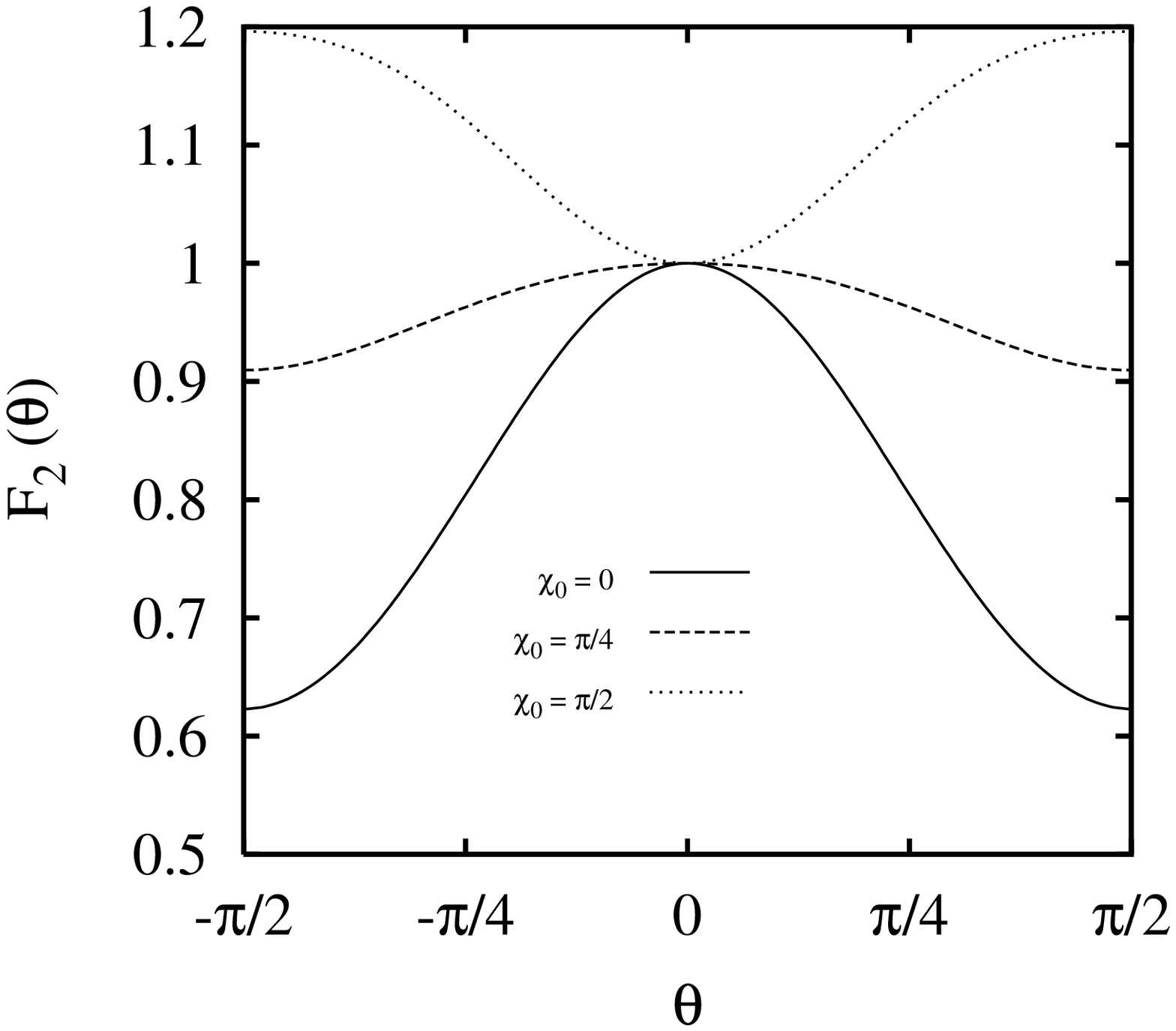,width=7cm,height=4.5cm}}
\caption{ $F_1(\theta)$ in Eq.(\ref{8}) and $F_2(\theta)$ in
Eq.(\ref{9}) are shown,  which represent the angle dependent
$\kappa_{yy}$
 in the superclean limit and clean limit, respectively.  }
\end{figure}
 Again the comparison  with the experimental  data suggests
 for  UPd$_2$Al$_3$. On the
other hand the microscopic model  $f=\cos(\chi)$ for UPd$_2$Al$_3$
appears to suggest $f=\cos(\chi)$\cite{13}.
 We are benefited  with useful  discussions with
Stephan Haas,  Yuji Matsuda,  Peter Thalmeier ad T. Watanabe on
UPd$_2$Al$_3$. \vspace{-0.4cm}

%
%
%
%

%
%
%
%


\end{document}